\documentclass[final,1p,times]{elsarticle} % do not change
\usepackage{graphicx} % do not change
\usepackage{epsfig}
\usepackage{amssymb} % do not change
\usepackage{amsthm} % do not change
\usepackage{lineno} % do not change

\journal{Nuclear Physics A} % do not change
\begin{document} % do not change

\begin{frontmatter} % do not change

%% QM09Author: please enter your  
%% Title, author and address info here; please do not use footnotes

% Your Title - please insert
\title{From $R_{AA}$ via correlations to jets - the long road to tomography}

% Principle author, and co-authors - please insert
\author{Thorsten Renk}

% Address - please insert
\address{Department of Physics, P.O. Box 35, 40014 University of Jyv\"{a}skyl\"{a}, Finland and\\
Helsinki Institute of Physics, P.O. Box 64, 00014 University of Helsinki, Finland}

\begin{abstract} % do not change
%% Text of abstract goes here - please insert

The main motivation to investigate hard probes in heavy ion collisions is to do tomography, i.e.
 to infer medium properties from the in-medium modification of hard processes. Yet while the suppression of high $P_T$ hadrons has been measured for some time, solid tomographic information is slow to emerge. This can be traced back to theoretical uncertainties and ambiguities in modelling both medium evolution and parton-medium interaction. Ways to overcome these difficulties are to constrain models better and to focus on more differential observables. Correlations of high $P_T$ hadrons offer non-trivial information beyond what can be deduced from  single hadron suppression. They reflect not only the hard reaction being modified by the medium, but also the back reaction of the medium to the hard probe. Models for hard back-to-back correlations are now very well constrained by a wealth of data and allow insights into the nature of the parton-medium interaction as well as first true tomographic results. Models of full in-medium jet evolution are being actively developed, but have yet to make substantial contact with data. Progress is slower in the understanding of low $P_T$ correlations, the ridge and the cone, although a qualitative understanding of the nature of the physics behind these correlations starts to emerge.

\end{abstract} % do not change

\end{frontmatter} % do not change

%% QM09: we keep linenumbers at least for initial version
%\linenumbers % do not change

%% start of main text - please insert. 

%%\section{}\label{}

\section{Tomography with hard probes}

In ultrarelativistic heavy-ion (A-A) collisions, a new state of hot and dense QCD matter is produced and offers the exciting prospect of studying collectivity and a phase transition in a quantum field theory experimentally.  The idea to utilize tomography as a tool to study the properties of this state of matter is rather simple: The basic principle of tomography is to generate a probe with known properties, propagate it through a medium and infer the properties of the medium from the changes induced on the probe. In A-A collisions, hard processes occuring along with the production of the medium generate suitable tomographic probes in terms of high $p_T$ partons. After their initial production with high virtuality, they undergo the quantum evolution towards lower virtuality scales  while interacting with the surrounding medium, the net result of which is the suppression of observed hard processes as compared to the scaled expectation from proton-proton (p-p) collisions.

In order to describe the physics underlying this suppression theoretically, three ingredients are needed. Fist, there is the hard process itself, which due to a separation of scales can be treated distinct from the medium dynamics and is calculable with good accuracy in perturbative QCD. Then, there is the interaction of the evolving parton shower with the medium. This however depends on the unknown, non-perturbative properties of the medium (e.g. in terms of relevant degrees of freedom) and ultimately needs to be inferred from measurement. Finally, there is the complication that the medium itself evolves at the same timescales set by the propagation of the hard probe, thus the evolution of the medium density as a function of space and time needs to be known. While there are good constraints from bulk matter observables, the details of this evolution are not known with certainty.

Tomography in the usual sense would be knowing the source and the parton-medium interaction and inferring the density evolution from those, However, since there are two unknowns, a different analysis strategy is needed. A promising way to proceed is to focus on all available constraints rather than a single high $P_T$ observable as often done. This, in short, is the reason for the importance of hard correlation measurements --- they offer a handle to observe the medium in a different way in terms of in-medium pathlength distribution and type of the high $p_T$ parton and hence allow ultimately together with other observables to separate the interaction dynamics between hard probe and medium from the evolution dynamics of the medium.

Conceptually, this requires to compute hard processes in a medium model constrained by bulk observables such as a hydrodynamical model. In this way, combinations of medium models and parton-medium interaction models can be tested against different observables. Thus, jet tomography can be a well constrained science if the full wealth of data provided by experiments is used. An example of such an approach is shown in \cite{Comparison} where three different models for the parton-medium interaction (AMY, ASW and HT) are used within the same hydrodynamical background to compute hard single hadron observables. It is clearly seen that while the numerical extraction of medium parameters leads to values of the transport coefficient $\hat{q}$ different up to a factor two, qualitatively the models are very similar in terms of the $P_T$ dependence of single hadron suppression, the extrapolation to different collision centrality or the way they probe geometry. To second order however, they predict somewhat different behaviour of the suppression with respect to the angle of the hard hadron with the reaction plane, a property which can be used for discrimination. In the following, it will be shown that correlation measurements are even more powerful to discriminate between models and hence to extract medium properties.

\section{Dihadrons and jets --- hard correlations}

Any hard process in QCD creates a back-to-back pair of partons at high virtuality which subsequently evolve through a partonic shower with decreasing virtuality into hadron jets. If one focuses on single hard hadron observables, one is biased towards events in which a single parton carries most of the momentum of the shower. The fragmentation function usually includes effects of both perturbative shower evolution and non-perturbative hadronization, but in this particular case it chiefly describes the hadronization of the leading parton. Therefore the approximation is often made that the medium effect can be cast into the form of energy loss to the medium, followed by vacuum fragmentation. The medium-modified fragmentation function for a parton with initial energy $E$ and virtuality $Q$ which describes the production of hadrons at momentum fraction $z$ from this parton can then be cast into the schematical form

\begin{equation}
D_{med}(z, Q) = P(\Delta E, E) \otimes D_{vac}(z, Q \rightarrow Q_h) \otimes D_{vac}(z, Q_h)
\end{equation}

where $P(E, \Delta E)$ is the probability distribution for energy loss $\Delta E$ in the medium given initial energy $E$ and $D_{vac}(z, Q \rightarrow Q_h)$ describes the perturbative partonic shower evolution decreasing the virtuality scale from $Q$ down to a hadronic scale $Q_h$ whereas $D_{vac}(z, Q_h)$ stands for the non-perturbative hadronization process.

The energy loss picture is applicable for observables which are sensitive to the leading hadron in a shower. This includes single hadron observables and also the hard back-to-back correlation with a trigger. However, in order to discuss correlations on the same side of the trigger, the approximation is not sufficient as it does not treat subleading partons correctly, and in the case of fully reconstructed jets where there is no bias towards a single parton carrying most of the energy, the energy loss picture is insufficient. Here, the whole parton shower needs to be evolved in the medium, whereas hadronization can, due to uncertainty principle arguments, still be assumed to take place outside the medium. The schematical expression for the medium-modified fragmentation function is then

\begin{equation}
D_{med}(z, Q) = D_{med}(z,Q \rightarrow Q_h) \otimes D_{vac}(z, Q_h).
\end{equation}

\subsection{Dihadron correlations and pathlength}

Hard back-to-back hadron correlations probe the leading hadron(s) of each shower and hence the energy loss picture can be applied. Calculations have been done in NLO pQCD for a hard sphere overlap with Bjorken expansion using the ZOWW energy loss formulation \cite{ZOWW}, using LO pQCD and the ASW energy loss formulation \cite{ASW} in 2-d \cite{Dihadrons1,Dihadrons2} and also as a function of the angle with the reaction plane in 3-d hydrodynamical expansions \cite{Dihadrons3}. A calculation with elastic energy loss in a 3d-hydrodynamics is presented in \cite{Elastic}. Thus, at least some of the systematics is known and it can be inferred what properties of the system are reflected in such correlations.

\begin{figure}
\begin{center}
\epsfig{file=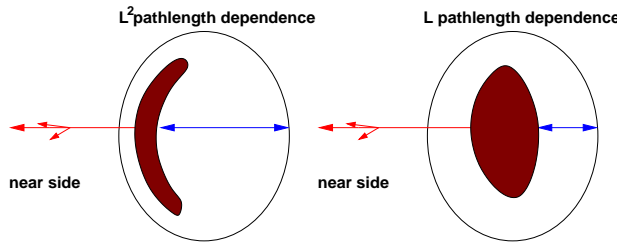, width=8cm}
\caption{\label{F-Pathbias}A schematic picture of the surface bias effect for different parametric pathlength dependence of energy loss.}
\end{center}
\end{figure}

It emerges from these studies that dihadron correlations are an excellent tool to probe differences in the parametric dependence of energy loss on pathlength. The reason is apparent from Fig.~\ref{F-Pathbias}: If one triggers on a hadron (red) on the near side, the most likely vertex of origin will be close to the surface of the medium, and this bias will be stronger for higher powers of the pathlength dependence of energy loss. However, this in turn influences the average away side pathlength (blue) which becomes larger for higher powers of pathlength dependence and moreover is weighted with that higher power. Therefore, a distinction between $L$-pathlength dependence (in a constant medium) as characteristic for elastic energy loss and $L^2$ dependence is very clear in back-to-back correlations and leads to factors two or more difference in observables \cite{Elastic}. A conclusion common to the ZOWW \cite{ZOWW} and the ASW computation of dihadron correlations \cite{Elastic} is that the average contribution of elastic energy loss is  $O$(10\%).

\begin{figure}
\begin{center}
\epsfig{file=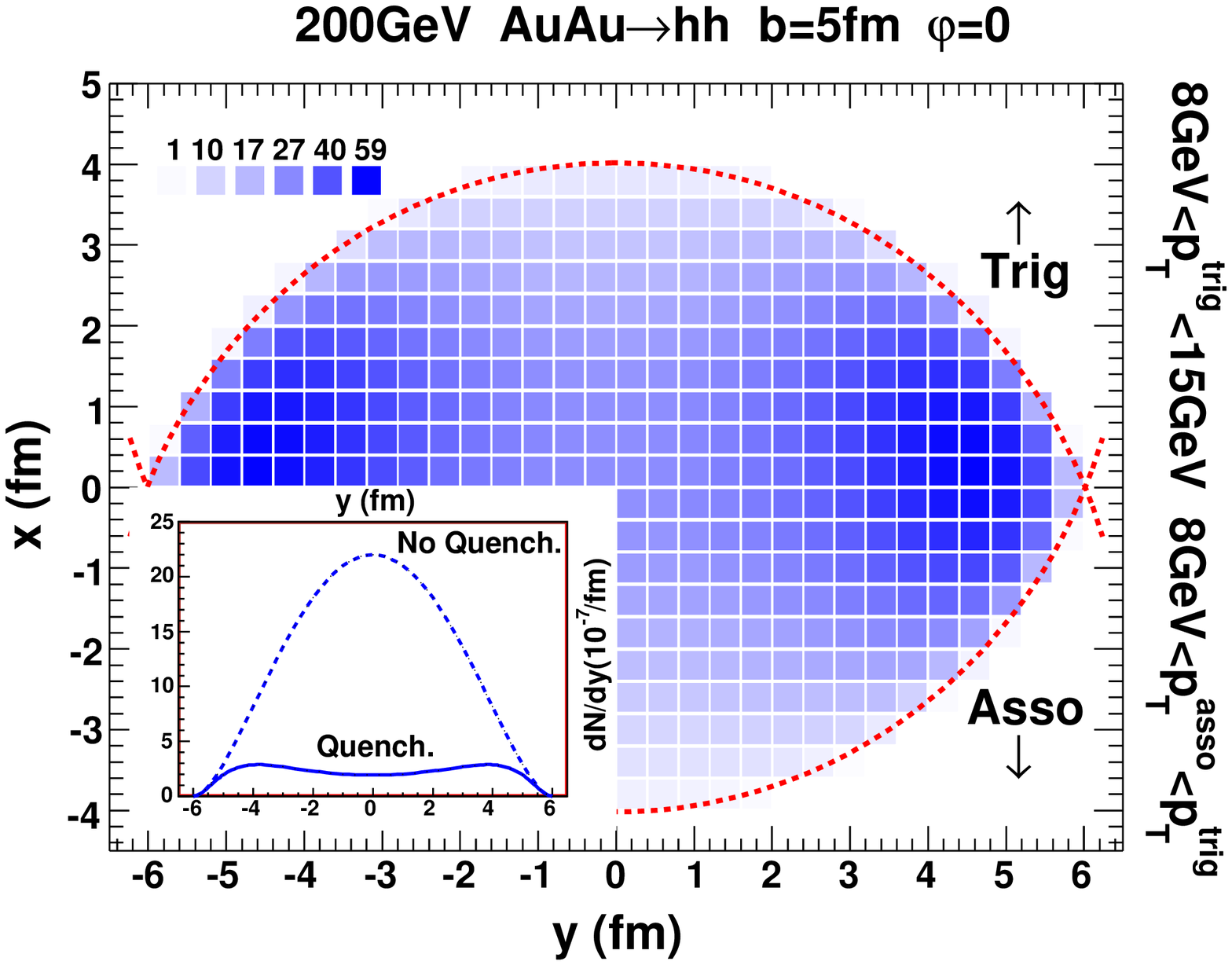, width=6cm}\epsfig{file=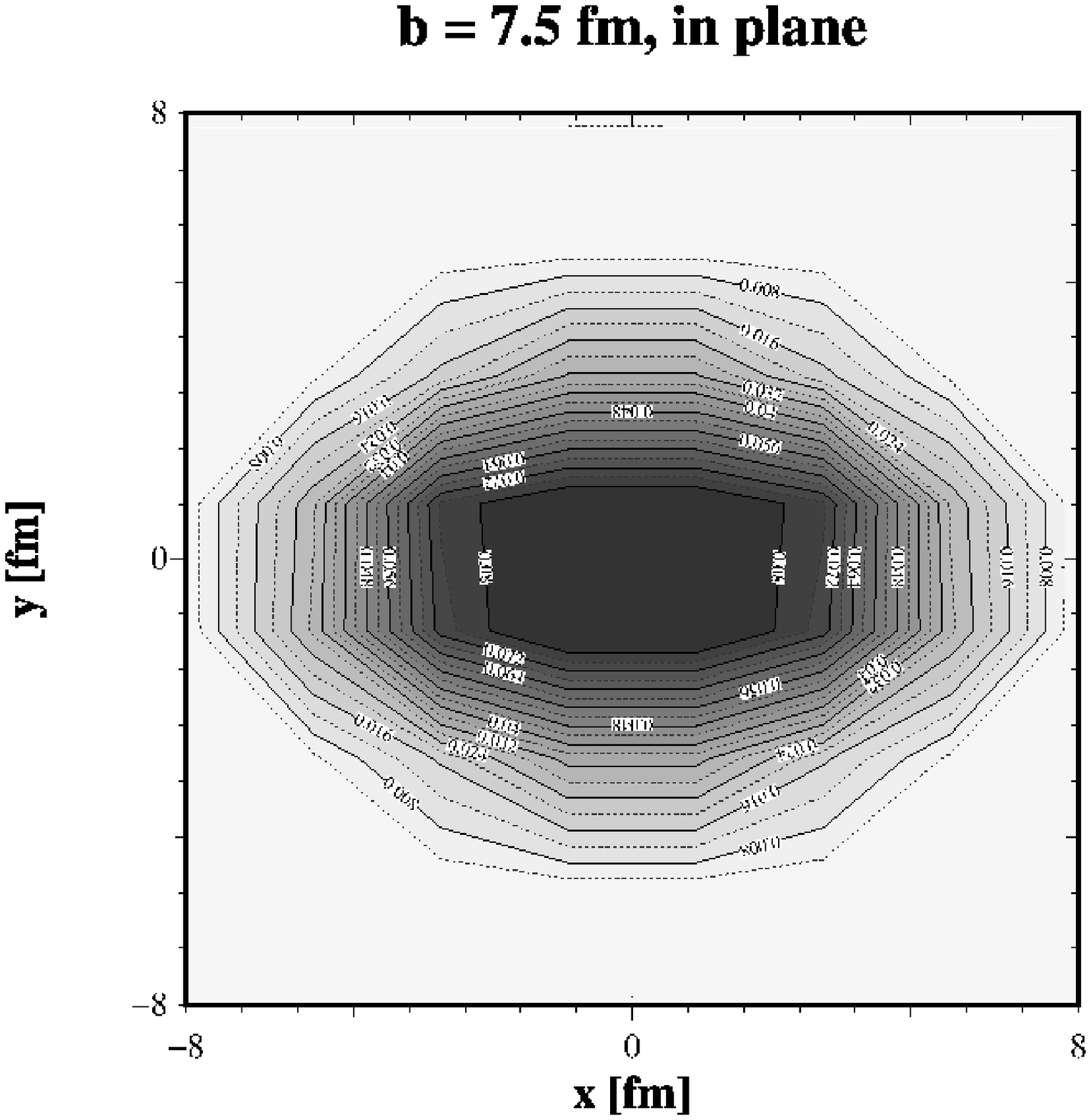, width=5cm}
\caption{\label{F-Geometry}The degree of tangential bias for dihadron correlations in the ZOWW formulation of energy loss \cite{ZOWW} and in the ASW formulation \cite{Dihadrons3}.}
\end{center}
\end{figure}

The formalisms differ however in details, such as the geometrical picture of the suppression. As shown in Fig.~\ref{F-Geometry}, the ZOWW computation shows a strong tangential bias whereas the ASW computation does not. Presumably the reason for this difference is that the ZOWW picture utilizes an averaged energy loss given a path through the medium whereas the ASW picture allows for fluctuations around this average, and as a result the correlation between position and energy loss is weakened. More differential observables, such as back-to-back correlations as a function of the angle with the reaction plane, which are directly sensitive to the degree of tangential bias, have been calculated \cite{Dihadrons3} and are currently being measured \cite{McCumber}. They may ultimately also probe the steepness of the initial medium density profile and allow to make a distinction between Glauber and CGC initial conditions in hydrodynamical evolution models and hence also have relevance for the discussion of the viscosity of the medium. If this comparison between theory and experiment can be made quantitatively, it would be the first real success of tomography in the sense of measuring a density distribution.

\subsection{Fully reconstructed jets}

If one is interested in studying correlations on the same side of a hard trigger hadron, one probes subleading shower hadrons. Continuing to correlate more and more particles down to lower momentum scales, one eventually includes all hadrons of a shower in the analysis and hence arrives at fully reconstructed jets. High $P_T$ physics with jets has a number of conceptual advantages over physics with single hard hadrons. First, jets in which the original shower initiator momentum is shared across many soft hadrons are frequent, those in which a single hard hadron takes most of the initial momentum are not. Thus, jet physics has higher statistics and can reach out to larger partonic $p_T$ than single hadron physics. Second, the shower evolution in the medium happens against a 'meter stick' and a 'clock' given by the medium extension and lifetime, which allows to probe the jet evolution not only in momentum space but also in position space. But most important, while different models in the energy loss picture may agree on the amount of energy lost from the leading parton, disucssing fully reconstructed jets also requires a picture how this lost energy is redistributed. In particular, different pictures of leading parton energy loss lead to characteristic effects in the distribution of subleading hadrons.
This was pointed out first in \cite{HBP} where it was argued that radiated gluons in a radiative energy loss picture lead to a charcteristic enhancement in the low $P_T$ region of the longitudinal momentum distribution inside medium-modified jets as compared to jets in vacuum.

Theoretically, the most promising approach to medium-modified jet physics are Monte-Carlo (MC) simulations. They build on well-tested vacuum shower evolution codes such as PYSHOW \cite{PYSHOW} or HERWIG \cite{HERWIG} under the assumption that only the partonic evolution of the shower is affected by the medium, but that hadronization happens sufficiently far outside the medium that it can be treated as in vacuum. Such MC codes have the advantage that they treat energy-momentum conservation explicitly at each vertex and furthermore they allow to analyze the output with the same techniques used to find jets in experimental data.
Currently, three major codes are being developed: JEWEL \cite{JEWEL1,JEWEL2} follows the philosophy to modify the interaction of shower and medium using perturbative QCD under the assumption that the medium constituents can be resolved as thermally distributed quarks and gluons. Q-PYTHIA and Q-HERWIG \cite{QPYTHIA1,QPYTHIA2} are the direct extension of the ASW formalism \cite{ASW} from leading parton energy loss to the evolution of a whole shower. Finally, YaJEM \cite{YAS1,YAS2,YAS3}  is based on modifications of parton kinematics in the shower where the medium effect is introduced in an {\em ad hoc} way in different scenarios (induced radiation and drag force) to avoid any {\em a priori} assumptions about the nature of the medium degrees of freedom.
Jet observables like the longitudinal distribution of hadron momenta in the jet, the thrust distribution or the angular distribution of hadrons in the shower have been obtained in all codes. However, the medium-modified shower picture should also account for observables sensitive to the leading shower hadron only. 

\begin{figure}
\begin{center}
\epsfig{file=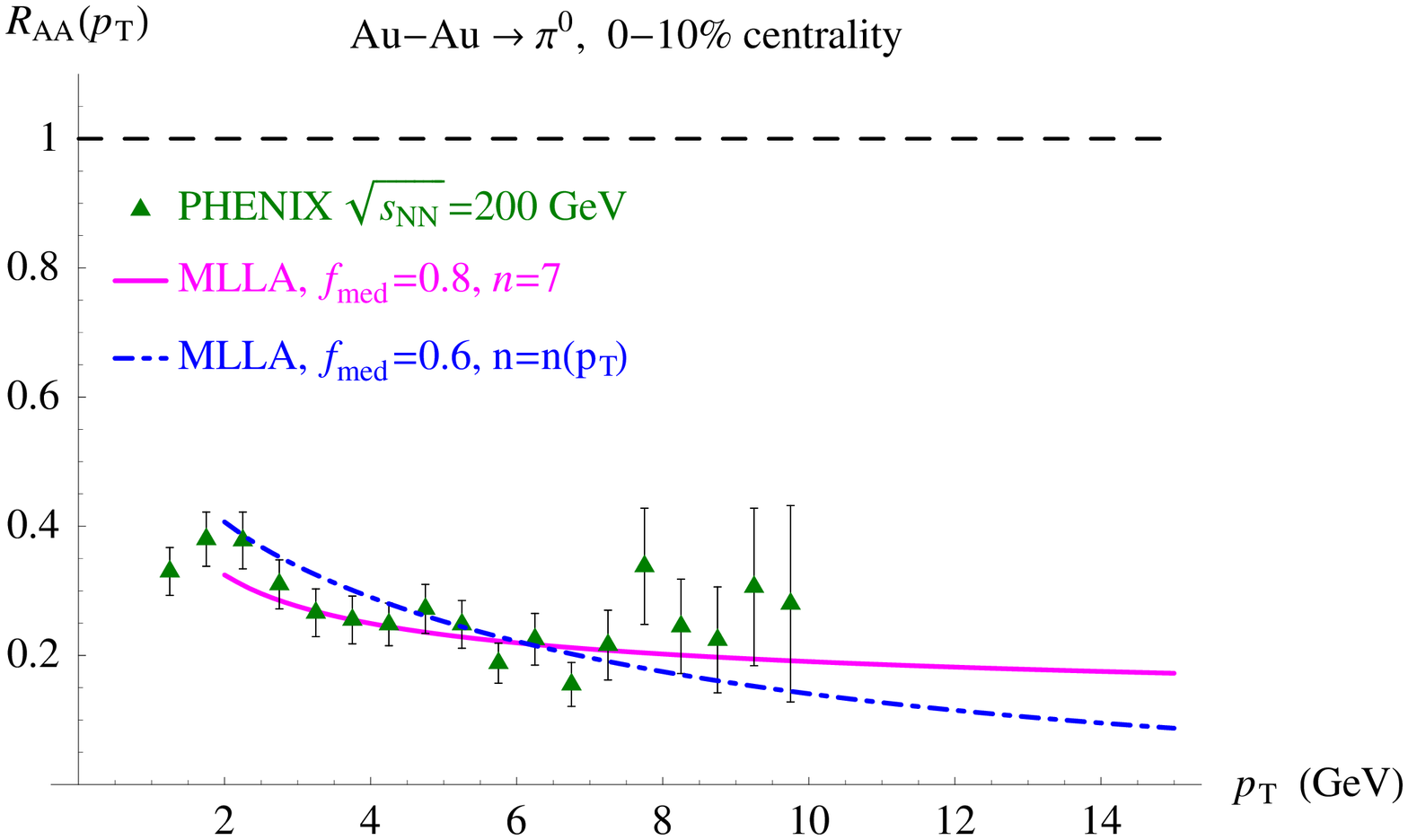, width=6cm}\epsfig{file=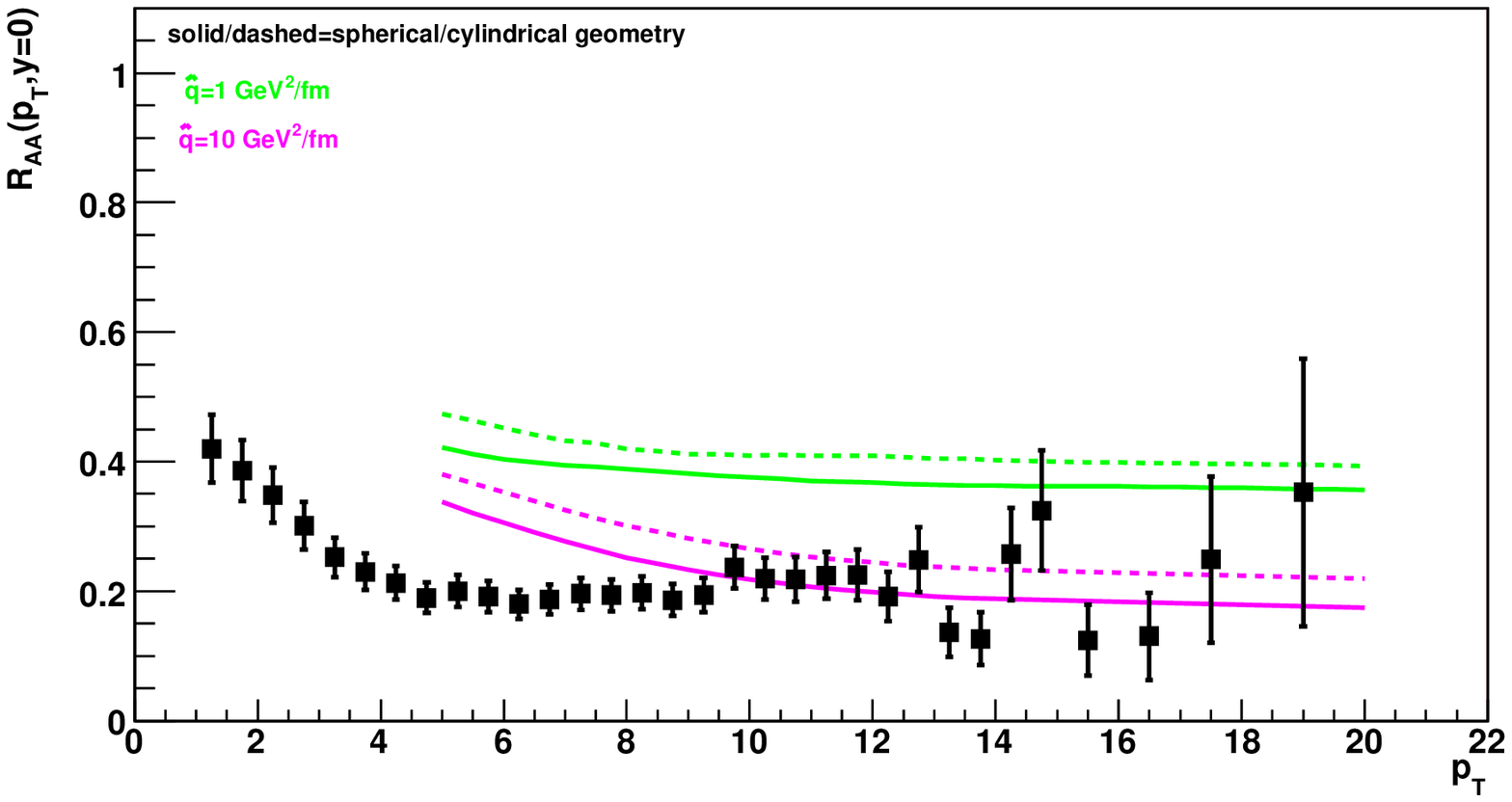, width=7cm}
\caption{\label{F-RAA}Nuclear suppression factor $R_{AA}$ computed for a medium-modified shower in an analytical approximation \cite{HBP} (left) and in Q-PYTHIA \cite{QPYTHIA1} (right) compared with PHENIX data \cite{PHENIX_R_AA}.}
\end{center}
\end{figure}

As an example, Fig.~\ref{F-RAA} shows the nuclear modification factor $R_{AA}$ computed in the analytical approximation of \cite{HBP} for a schematic treatment of the medium and in Q-PYTHIA averaged over a cylindrical medium geometry \cite{QPYTHIA1}. Similar results have also been obtained in YaJEM averaged over a 3-d hydrodynamical model \cite{YAS1,YAS3}. In all cases, a falling trend of $R_{AA}$ with $P_T$ is observed, regardless of the averaging over the medium. At present this is not supported by the trend of the data which seems to be increasing, although errors are substantial. However, should the increasing trend of the data be confirmed by a higher statistics analysis, it would indicate a conceptual issue which is not properly understood. In a similar line, a detailed investigation of the LPM effect in the MC \cite{JEWEL2} indicates that the $L^2$ dependence of radiative energy loss is only visible for a certain set of assumptions. If these are relaxed to a more realistic scenario, the $L^2$ dependence is hardly visible. However, phenomenologically this poses a problem, as hard back-to-back correlations strongly favour a parametric $L^2$ dependence of energy loss. It appears necessary at this point to test all in-medium shower codes thoroughly against existing leading hadron observables before they are applied to predict jet physics.

\section{The Cone and the Ridge --- medium recoil}

While in-medium shower evolution codes treat the production of soft partons perturbatively and assume their hadronization can be described as in vacuum, there is no reason to assume that this corresponds to the actual dynamics of a jet in the medium at low $P_T$ which is in all likelihood non-perturbative. Instead, observed correlations of low $P_T$ hadrons with a hard trigger may rather reflect the way the medium responds (in terms of recoil or shockwaves) to the energy deposition by a propagating jet. Thus, more phenomenological models are currently used in attempts to describe the soft correlations on the same side and the away side of a trigger hadron.

In discussing the general structure of the correlation pattern, it is useful to keep a few things in mind. If one triggers on a high $P_T$ hadron on the near side, there must be a correlation on the away side due to momentum conservation. The interesting physics is only in the dependence of this away side correlation on the azimuth $\phi$, the rapidity $\eta$ and $P_T$, in other words in the precise way energy and momentum are balanced. In addition, other physics may lead to a correlation on the near side, for example the leading hadron in a jet is correlated with subleading hadrons, but this correlation is not required by momentum conservation but arises due to QCD. In a medium, other phenomena may also be acidentially correlated with a hard hadron. For example, hard hadrons tend to emerge perpendicular to the medium surface because this minimizes the in-medium path and hence energy loss. At the same time, radial flow is perpendicular to the surface because this is the direction of the pressure gradients. But although this correlates flow with the trigger, the two phenomena are unrelated in the sense that radial flow is in no way be caused by the hard hadron. Finally, any near side phenomenon correlated with the trigger (be it an accidential correlation or not) will also cause a correlation on the away side due to momentum conservation. All these effects need to be disentangled carefully.

\subsection{The Ridge}

The most prominent near side correlation associated with a trigger is the so-called ridge, a structure which is visible at low $P_T$, extends out to at least 4 units of rapidity in $\eta$ with an almost unchanged width in $\phi$ \cite{PHOBOS-Ridge}. These features practically rule out any connection of the ridge correlation with energy loss. The ridge cannot contain radiated quanta, as $R_{AA}$ places a constraint on the average amount of energy lost into the medium. A number for the lost energy, given that a trigger was observed, has been obtained e.g. in the formalism of \cite{EDEP} and is $O(0.5)$ GeV. This is insufficient to explain the strong hadron production out to 4 units in $\eta$ which takes a substantial amount of energy. An alternative explanation of the ridge as a medium recoil due to energy lost into the medium fails, since due to the coherence time effect, energy deposition into the medium happens comparatively late, but causality requires large spacetime rapidity correlations to arise early. However, if one invokes medium recoil, the medium spacetime rapidity must equal its momentum rapidity due to the observed Bjorken flow of the medium. Thus, the ridge cannot be interpreted as a medium recoil. Besides, a ridge is also observed in untriggered correlations, indicating that its correlation with a hard trigger is accidential. On the other hand, the properties of the ridge seem well accounted for in models which view the ridge as an initial state such as the flux tube model \cite{Flux1,Flux2}.

\subsection{The Cone}

The away side correlation at low $P_T$ takes the form of a broad double-hump structure in $\phi$ (its structure in $\eta$ is not easily observed as the rapidity of the away side parton is only weekly constrained by kinematics). It is most commonly interpreted as conical emission. Early models to explain the cone in terms of Cherenkov radiation \cite{Cherenkov} predicted a $P_T$ dependence of the cone opening angle which was not found in the data. Therefore the current theory efforts to describe the cone focus on the hydrodynamical response to energy deposition into the medium.

The main conceptual issue is finding the source term for the hydrodynamical evolution which describes the physics of energy loss into the medium adequately.  Such efforts include a Bethe-Bloch source term \cite{Source1}, a source term based on ASW energy loss \cite{Source2}, or the Higher Twist formulation of energy loss \cite{Source3} and also ideas utilizing the AdS/CFT correspondence which descibes the physics of a gauge theory in the strong coupling limit \cite{AdS1,AdS2}.

A problem of the theoretical descriptions of the cone is that many different scenarios manage to result in a double-hump structure of the away side correlation pattern, for example bulk statistical momentum conservation and elliptic flow \cite{Borghini}, an energy loss source term and hydrodynamical evolution \cite{Source2}, recombination in a linearized hydrodynamical model \cite{Torrieri} and the recoil of the bulk medium to an initial state flux tube correlation \cite{Rio}. This, and the observation that the observed angle is not even different at SPS energies makes it likely that the observed double-hump structure does not reflect details of the dynamics, but is rather driven by a combination of energy-loss caused surface bias, a position-flow correlation in hydrodynamical models and a kinematic bias towards alignment of radial flow and large angle correlation, relatively independent of the underlying dynamics as long as there is a generic transport of energy and momentum to large angles. Such a scenario has been explored in \cite{Mach1,Mach2,Mach3} and found qualitative agreement with the data on 2-particle correlations. There are also measurements of 3-particle correlations, however here neither the experimental analysis nor the theoretical calculation are conclusive yet.

\section{Conclusions}

In discussing correlations with respect to a hard hadron trigger as a tomographic tool, it is useful to make a distinction between hard and soft associate hadrons. The reason is that due to the presence of a hard scale, the virtuality, a hard process in medium and the subsequent partonic evolution of a shower remain governed by perturbative QCD dynamics, even if the medium itself and its effect on the shower evolution is not. This argument, however, does not apply to soft correlations in the typical momentum range of the bulk --- here, one cannot maintain an easy distinction between recoiling medium and modified shower. 

In the hard sector which can be described in the leading parton energy loss picture, highly constrained calculations are possible and being done and begin to answer questions about the medium. In practice, this requires to give up schematic descriptions of the medium (Bjorken cylinders) and include all constraints of bulk observables by modelling the medium with e.g. state of the art hydrodynamical models, and in addition to apply a model not only to single hadron supression but also to more differential hard observables, including correlations. Such calculations have established reasonably well that elastic energy loss is a small contribution in the light quark sector and that the core of the medium is not completely black. On the experimental side, more differential high $P_T$ observables are reaching the precision necessary to distinguish between models.
For observables sensitive to subleading hadrons in the shower, and eventually fully reconstruced jets, MC codes for in-medium parton showers are currently being developed and tested. They investigate various different ideas how to implement the effect of the medium. However, for most codes no thorough comparison with leading hadron observables has been done yet, and there are hints of unresolved issues, such as the decreasing trend of $R_{AA}$ with $P_T$ or the almost linear growth of energy loss with pathlength.

Soft correlations have to be modelled in more phenomenological approaches. There is mounting evidence that the near side ridge correlation is not directly connected with energy loss. However, on the away side, due to momentum conservation it is quite clear that at least part of the cone correlation must have such a connection. At present, the data favour models describing the cone as a recoil of the bulk medium over those describing it by a medium modified fragmentation process. Various ideas for the recoil dynamics are explored, however there are indications that the observed correlation is mainly driven by a combination of different biases rather than the underlying dynamics. Despite good experimental data, a consistent theory of high $P_T$ bias and flow-modified medium recoil is very complicated and at present only capable of qualitative answers. Any claims that one would already have measured medium properties such as the speed of sound appear hence premature.

%% end of main text

\section*{Acknowledgments} 

This work was financially supported by the Academy of Finland, Project 115262. 

% please insert, comment out or delete if not needed
%This is where one places acknowledgments for funding bodies etc., if needed.
%For the large collaborations, this is listed once and for all, together with 
%the author lists etc. in the proceedings back-material.

 % do not change 
\end{document}